\documentstyle[editedvolume]{crckapb} 

\begin{opening}
\title{THE LITHIUM PLATEAU ENIGMA
}

\author{C.Charbonnel}
\author{S.Vauclair}
\institute{Laboratoire d'Astrophysique de Toulouse - OMP\\
           14, av.E.Belin, 31400 Toulouse, France}
\end{opening}

\runningtitle{THE LITHIUM PLATEAU ENIGMA}

\begin{document}

\section{The lithium plateau enigma may be summarized as follows : }

Main-sequence Pop II stars stars with effective
temperatures between 5500K and 6500K show a remarquably constant value
for the lithium abundance ($\Longleftrightarrow$ the lithium plateau).
Furthermore the dispersion around this value is
very small (Bonifacio \& Molaro 1997). 
However, from the observations of Pop I stars, there is strong evidence that
the lithium abundance highly varies from star to star.
The variation with T$_{\rm eff}$ and age clearly appears in the galactic
cluster data. 

From the theory and modelisation of stellar internal structure,
lithium is expected to vary from star to star due to both nuclear destruction
and/or element settling. These effects account well for the
observations of Pop I stars, although more quantitative comparisons between
observational and theoretical results are still needed.
Helioseimology now provides a spectacular confirmation of the precision
we have attained in the modelisation of the solar internal structure, including
element settling (Richard et al. 1996). 

So why is the lithium abundance constant in the so-called lithium
plateau while all predictions suggest that it should vary from star to star?
Is there an ``abundance attractor" which would work in Pop II stars but not in
Pop I stars?

\section{Hints for a solution}

Several models have been proposed to account for the lithium plateau.
The ``old standard model" in which no settling was introduced is excluded as
unphysical. In the ``mass loss model" (Vauclair \& Charbonnel 1995),
a stellar wind is supposed to prevent element settling during the
stellar lifetime. In the ``rotation model" (Pinsonneault et al. 1990, 
Charbonnel \& Vauclair 1999), the Pop II stars are supposed to have been
mildly mixed below the convection zone due to rotation-induced shears.
In any case the solution seems somewhat ``ad hoc" as it assumes
that some parameters are fixed in all stars (the mass loss rate or the
rotation rate) for the lithium value to remain constant along the plateau.
It would be much more satisfying to find a ``lithium abundance
attractor" which would remain stable in halo stars while
fundamental parameters (M$_*$, T$_{\rm eff}$, [Fe/H]) vary.

\section{Lithium abundance attractor}
Such an attractor may exist (Vauclair \& Charbonnel 1998) :
Indeed, the lithium profiles inside the Pop II standard stellar models
including element segregation present a maximum value, Li$_{\rm max}$,
which remains constant all over the range in T$_{\rm eff}$ and metallicity
of the plateau while the surface value is expected to change. 

This result leads to the idea that the observed lithium abundances may be
related to Li$_{\rm max}$. Since the observations of the lithium in the 
plateau reveal a very small dispersion around a stable value, this value 
must indeed lie close to Li$_{\rm max}$. 
In this case the derived primordial value is 2.35. 
When compared to BBN computations (Copi et al. 1995)
this result leads to a baryonic number between 1.2 and 5 $10^{-10}$.
For H~=~50, this value corresponds to $0.018 < \Omega _{b} < 0.075$.
The macroscopic process which would act in the way of moving this lithium
up to the surface is still to be found.

{}

\end{document}